\documentclass[conference]{IEEEtran}
\IEEEoverridecommandlockouts
\usepackage{cite}
\usepackage{amsmath,amssymb,amsfonts}
\usepackage{algorithmic}
\usepackage{graphicx}
\usepackage{float}
\usepackage{textcomp}
\usepackage{xcolor}
\def\BibTeX{{\rm B\kern-.05em{\sc i\kern-.025em b}\kern-.08em
    T\kern-.1667em\lower.7ex\hbox{E}\kern-.125emX}}
\begin{document}

\title{Heart Rate Classification in ECG Signals Using Machine Learning and Deep Learning}

\author{\IEEEauthorblockN{\textsuperscript{} Thien Nhan Vo}
\IEEEauthorblockA{\textit{Institute of Engineering} \\
\textit{Ho Chi Minh City University of Technology (HUTECH)}\\
475A Dien Bien Phu, Ward 25  Binh Thanh District \\ Ho Chi Minh City \\ 
Email: thiennhan.math@gmail.com}

}

\maketitle

\begin{abstract}
This study addresses the classification of heartbeats from ECG signals through two distinct approaches: traditional machine learning utilizing hand-crafted features and deep learning via transformed images of ECG beats. The dataset underwent preprocessing steps, including downsampling, filtering, and normalization, to ensure consistency and relevance for subsequent analysis. In the first approach, features such as heart rate variability (HRV), mean, variance, and RR intervals were extracted to train various classifiers, including SVM, Random Forest, AdaBoost, LSTM, Bi-directional LSTM, and LightGBM. The second approach involved transforming ECG signals into images using Gramian Angular Field (GAF), Markov Transition Field (MTF), and Recurrence Plots (RP), with these images subsequently classified using CNN architectures like VGG and Inception.

Experimental results demonstrate that the LightGBM model achieved the highest performance, with an accuracy of 99\% and an F1 score of 0.94, outperforming the image-based CNN approach (F1 score of 0.85). Models such as SVM and AdaBoost yielded significantly lower scores, indicating limited suitability for this task. The findings underscore the superior ability of hand-crafted features to capture temporal and morphological variations in ECG signals compared to image-based representations of individual beats. Future investigations may benefit from incorporating multi-lead ECG signals and temporal dependencies across successive beats to enhance classification accuracy further.
\end{abstract}

\begin{IEEEkeywords}
Machine learning, Classification, ECG, Deep Learning
\end{IEEEkeywords}

\section{Introduction}
Heart arrhythmia is a condition characterized by irregular heartbeats due to disrupted electrical impulses within the heart \cite{ref3}. While some arrhythmias are benign, others can lead to severe health complications and even sudden cardiac death \cite{ref1}. The electrocardiogram (ECG) serves as a fundamental tool in diagnosing these irregularities, enabling the analysis of the heart's electrical activity over time \cite{ref4}. However, manual interpretation of ECG signals can be time-consuming and prone to variability, underscoring the need for automated and accurate classification methods.

Recent advances in machine learning (ML) and deep learning (DL) have shown remarkable success in ECG heartbeat classification tasks. For instance, deep convolutional neural networks (CNNs) have been leveraged to automatically extract morphological and temporal features from ECG signals, yielding impressive performance metrics in arrhythmia detection \cite{ref2}, \cite{ref4}. Furthermore, studies have proposed patient-specific classifiers that adapt to individual ECG morphologies, enhancing classification reliability \cite{ref3}.

In parallel, traditional ML approaches employing hand-crafted features—such as heart rate variability (HRV) and RR intervals—have demonstrated their efficacy in capturing clinically significant patterns in ECG data \cite{ref6}, \cite{ref7}. Notably, the combination of ML algorithms (e.g., LightGBM, Random Forest) with these features can outperform deep learning models, especially in scenarios with limited data or noisy signals \cite{ref1}, \cite{ref7}.

To explore the capabilities and trade-offs of these methodologies, this study investigates two primary approaches for ECG heartbeat classification: (i) traditional ML models trained on hand-crafted features and (ii) deep learning-based models using image representations of ECG signals, including Gramian Angular Field (GAF), Markov Transition Field (MTF), and Recurrence Plots (RP). By comparing these strategies, we aim to highlight the strengths and limitations of each technique in classifying arrhythmic and normal heartbeats, and to identify future directions for improving the accuracy and generalizability of automated ECG analysis.
\section{Materials And Methods}
\subsection{Preprocessing}
In the beginning, we decided to explore the given dataset and perform some basic pre-processing techniques. Each sig- nal from our dataset has three files where one contains the 2- lead ECG signal and the other two are representing the R-peak positions and the beat annotations. The first step in the pre- processing was to downsample the signal with higher sampling frequency, i.e 250 Hz so as to unify all the signals to the same sampling frequency (180 Hz). The second step was to filter the signal keeping only the frequency range between 0.5 and 35 Hz. Since our problem was to label ECG beats, we decided to split the signals into fixed-size beats, each containing 70 samples i.e. 35 samples before and after the R peak location. The next pre-processing step was to normalize all the beats so that they have the amplitude between -1 and 1. \cite{ref8} Finally, we had the dataset that contains separated beats and their labels.

\subsection{1st Approach}
Based on the feature extractions approaches in literature we added additional hand-crafted features to the data we fed to the classifier. We suspected that the average HRV, its median and variance, along with average signal amplitude might provide useful information to the model and the first tests confirmed the assumptions. Later, two more features (logarithm of the

current and the next RR distance) were added [8]. All of the aforementioned features were then concatenated with the 70- sample beats.

\begin{figure}[H]
    \centering
    \includegraphics[width=\linewidth]{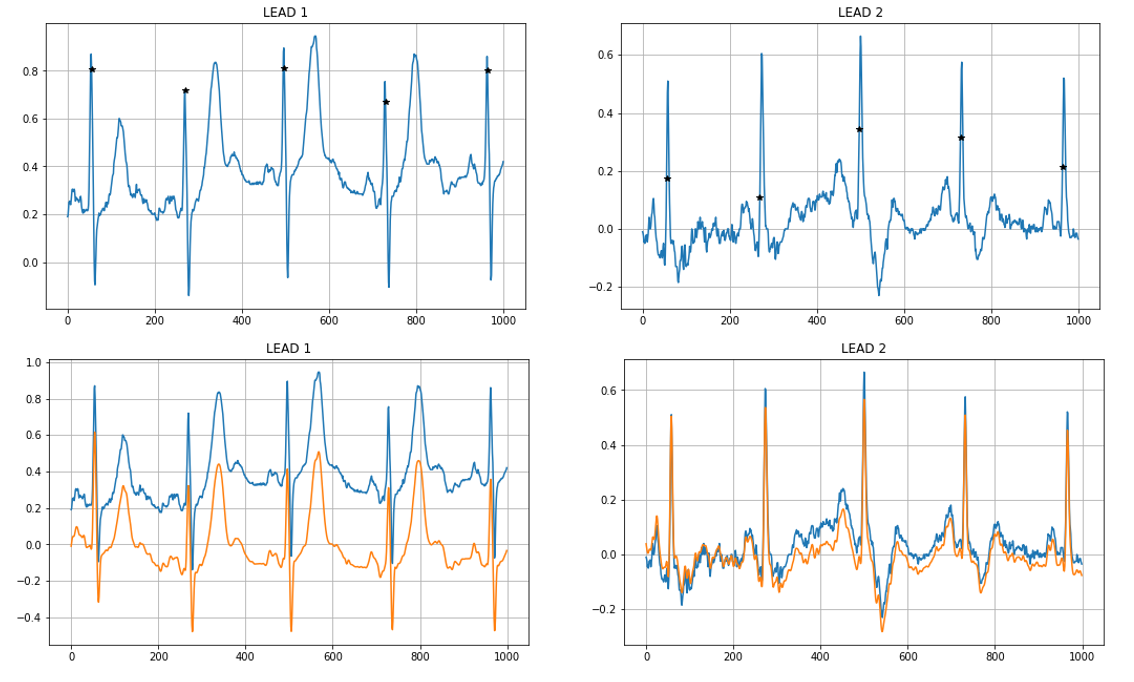}
    \caption{Original signal from both leads together with R-peak annotations and the signal after pre-processing steps}
    \label{fig:1}
\end{figure}

Having the pre-processed dataset, we started by researching existing approaches to the task and decided on two very different approaches. The main idea was to test both of them and to choose the one that would be more suitable for our classification problem.

One of the problems that we met here was the the un- balanced dataset. Namely, the majority of beats were labeled as normal beat, i.e. classified with the label ’N’. Therefore, we agreed on using the SMOTE technique to prevent this, instead of cutting of the huge amount of normal beats. The SMOTE is an oversampling technique where the synthetic samples are generated for the minority class. We decided to do the oversampling so that we have 100k ventricular (V) and super-ventricular (S) beats, and 300k normal (N) beats. It was applied only to the training data, while the test data remained unchanged so that it correctly represents the original data.

\begin{figure}[H]
    \centering
    \includegraphics[width=\linewidth]{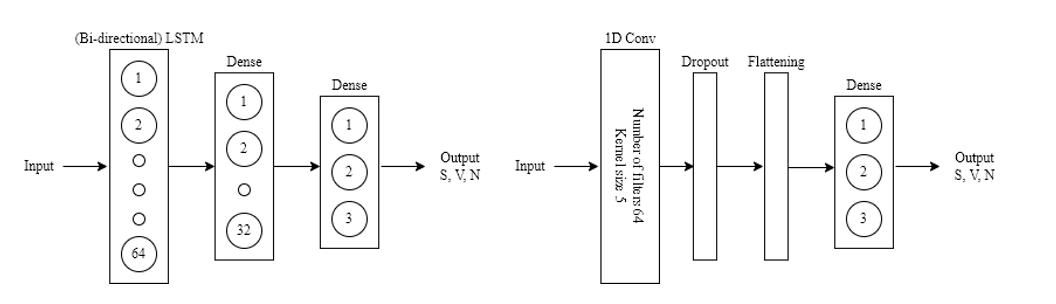}
    \caption{Models with LSTM, Bi-directional LSTM and 1D Convolutional layers respectively}
    \label{fig:2}
\end{figure}

After a thorough investigation of the dataset, we developed models capable of classifying the beats based on the input data. In the first approach, we tested the SVM, Random Forest, AdaBoost, Convolutional Neural Network using 1D convolutional layers, Recurrent Neural Networks using LSTM and Bi-directional LSTM layers, and finally, the ones that produced the best results, LightGBM models. All the model parameters were fine-tuned using GridSearch with a 3-fold cross-validation, which allowed us to quickly verify in which direction to move in when developing the classifiers.

\subsection{2nd Approach  Images from ECG}

The main idea is to exploit the state-of-the-art performances of 2d CNN in our classification problem.We need to transform each heartbeat in an image \cite{ref6}. We want a way to encode the temporal information of our signal into the processed images. To obtain our 2d matrix picture, three different methodologies were used: Gramian angular field (GAF), Markov transition field (MTF), and recurrence plots (RP). All of these methods were proven to be effective on time series regardless of the specific domain \cite{ref9}. I will briefly describe each methodology.

In GAF we change the representation of our signal by using the Polar Coordinate system. After normalization, we calculate the phase of each value in our time series using the following equation

\begin{equation}
\begin{cases}
\phi_i = \arccos(\tilde{x}_i) \\
r_i = \frac{t_i}{T},\ t_i \in \mathbb{N}
\end{cases}
\end{equation}

Where $x_i$ is the normalized value of a single data point in our time series. $t_i$ is the timestamp of that data point. $T$ is constant factor.

The final matrix contains the sum of the cosine of the 2 angles.
\[
\begin{pmatrix}
\cos(\phi_1 + \phi_1) & \cos(\phi_1 + \phi_2) & \cdots & \cos(\phi_1 + \phi_n) \\
\cos(\phi_2 + \phi_1) & \cos(\phi_2 + \phi_2) & \cdots & \cos(\phi_2 + \phi_n) \\
\vdots & \vdots & \ddots & \vdots \\
\cos(\phi_n + \phi_1) & \cos(\phi_n + \phi_2) & \cdots & \cos(\phi_n + \phi_n)
\end{pmatrix}
\]

It is a bijective function so we can recover the signal from the image. It represents the temporal correlation between the values of the time-series data. It maintains the temporal dependency.

In MTF an image represents a field of transition probabilities for a discretized time series. It's useful for representing the dynamics of the time series. The high-level idea is to divide the values of our data $\{x_1, ..., x_{tn}\}$ into $N$ quantile bins. From this, we can easily compute the Markov transition matrix. This matrix contains the probabilities of going from one quantile bin $i$ to another quantile bin $j$. The assumption is to consider our time series as a Markov chain and compute the frequencies. This matrix does not preserve the temporal dependencies and for this reason, we need to perform another transformation that creates the Markov transition field. The difference is that we create and put data on the quantile bins. Considering the exact timestamp(at k-th timestamp). In this case, we will preserve the temporal dependency by having the probability of going from bin i of timestamp k to bin j of timestamp l. We cannot recover the original signal given that we have a matrix with probabilities.

RP shows the collection of pairs of times at which the values of our time series have the same value(or similar values).
They are a graphical representation of the matrix

\[
R_{i,j} = \theta(\epsilon - \|x_i - x_j\|)
\]

where \( x_i \) stands for the value at time \( i \), \( \epsilon \) is a predefined threshold and \( \theta \) is the Heaviside function. One assigns a ``black'' dot to the value one and a ``white'' dot to the value zero. In our case we perform the recurrence plot even without using a predefined threshold.

In figure 3 we see all the heartbeat classes along with each transformation. All the transformations were achieved by using the Python library \texttt{pyts}. All the transformations produced images of 32x32 pixels. The choice is done given that the maximum number of pixels that we can have is 70 (70 is the total of timestamps in our signal).

\begin{figure}[H]
    \centering
    \includegraphics[width=\linewidth]{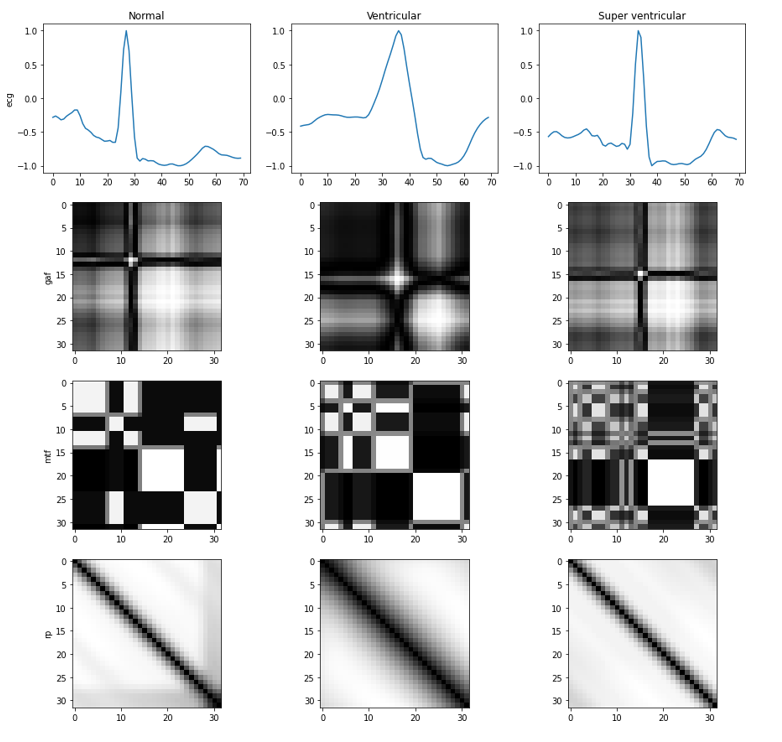}
    \caption{ECG image transformation}
    \label{fig:3}
\end{figure}

Every transformation has its own peculiarities and it focuses on different parts of the signal. The idea of using all of them comes from the fact that multi-modal architectures often performs better with respect to using a single modality. In our case we concatenate them in order to have as input a 3 channel image (the standard representation for rgb images). We use a convolutional neural network to extract the features and classify each beat. Different architectures were tried such as VGG, ResNet, Inception like architectures. The best model was a custom Inception base CNN that has 5 million parameters. Very similar performances were also achieved by simple VGG like architectures. The following parameters were used in the best configuration. We set the learning rate to be 03e-4, with batch size of 512. The loss function chosen was the categorical cross entropy. The dataset used is the same described in the previous chapter with the exception that no hand crafted feature was used, just the original ECG of the single heartbeat. The input are the images representing the single heartbeats.

\begin{figure}[H]
    \centering
    \includegraphics[width=\linewidth]{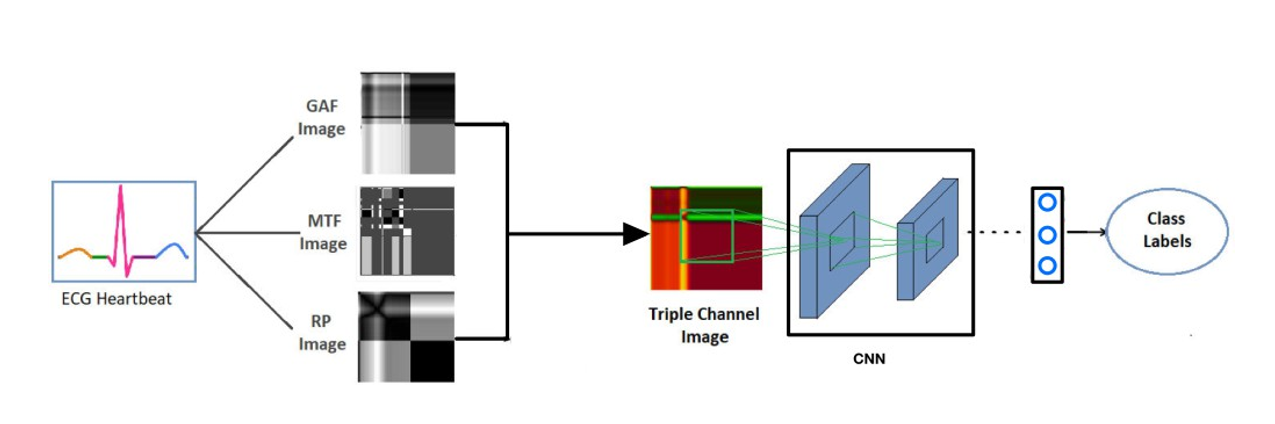}
    \caption{Architecture}
    \label{fig:4}
\end{figure}

\section{Results}
Below, we report a selected number of metrics that we observed in training the models.The metrics were the basis for the final model choice. The reported values are the macro averages of the metrics per label.
\begin{table}[h!]
\centering
\begin{tabular}{|l|c|c|c|c|}
\hline
\textbf{Model} & \textbf{Precision} & \textbf{Recall} & \textbf{Accuracy} & \textbf{F1 score} \\
\hline
Images approach       & 0.83 & 0.87 & 0.97 & 0.85 \\
Random Forest         & 0.93 & 0.93 & 0.99 & 0.93 \\
SVM                   & 0.59 & 0.58 & 0.58 & 0.58 \\
LightGBM              & 0.94 & 0.94 & 0.99 & 0.94 \\
LSTM                  & 0.86 & 0.95 & 0.98 & 0.90 \\
AdaBoost              & 0.43 & 0.72 & 0.70 & 0.44 \\
Bi-directional LSTM   & 0.91 & 0.94 & 0.98 & 0.93 \\
1D CNN                & 0.43 & 0.39 & 0.92 & 0.39 \\
\hline
\end{tabular}
\vspace{2mm}
\caption{Performance comparison of different models}
\end{table}
\section{ Discussion}
Based on the model scores, we immediately discarded the SVM, feed-forward neural network, and AdaBoost models. Their performances could have possibly been further im- proved, however, they were significantly lower than other models’, so we preferred to focus on the more powerful ones. Furthermore, the SVM model took a long time to train (approximately 4 hours), which meant investing more resources into an ineffective training. We were very satisfied with the performance of the Random Forest and LightGBM models. They are both ensemble techniques, which train smaller decision trees and then use either bagging (Random Forest) or boosting (LightGBM) to classify the samples. The advantage of the LightGBM is its fast optimization algorithm, which allowed us to train the model in approximately 6 minutes, leaving us more time to run a larger number of experiments.

The parameter which improved the results the most in the LightGBM model, was surprisingly the learning rate. By increasing it, one would normally risk not minimizing the gradient and reaching the minimum. However, a model with a relatively high learning rate was able to classify even the marginal labels better, possibly because of high penalization rates of L1 and L2 regularization. The ratio between under- sampling and oversampling also proved to be of value when trying to increase the scores.

Concatenating the two signal leads could further improve the results and add additional information to feed into the model. Research suggests that using the logarithms of a larger number of features might also render them more understand- able for the models. Furthermore, we discussed creating an LSTM model, which would take into account 4 previous beats and classify the fifth one, possibly finding the relationships between subsequent beats. Due to the limited time, we were unable to explore these approaches now, but it would be interesting to research it further in the future.

\section{Conclusion}
In the end, the model that performed best was the Light- GBM with the following parameters: learning rate of 0.5, max depth of trees 10, using multiclass logloss as metric, minimum data in leaves of 10, and 1000 estimators. We also set the alpha regularization to 0.5 and lambda regularization to 0.7327. The imbalanced dataset was handled using the SMOTE technique, undersampling the signal 300000 times and then oversampling it 100000 times.

We believe that the first approach worked best for the task because the hand-crafted features were able to capture the whole signal variation, whereas the images created contain information only on the single beat.

\end{document}